\documentstyle[preprint,aps]{revtex}
\def\be{\begin{equation}}
\def\ee{\end{equation}}
\def\bea{\begin{eqnarray}}
\def\eea{\end{eqnarray}}

\begin{document}
\draft

\voffset -0.1in
\title{SCOZA for Monolayer Films}
\author{Chi-Lun Lee$^1$ and George Stell$^2$}
\address{$^1$Department of Physics, State University of New York
  at Stony Brook, Stony Brook, New York 11794-3800 \\ $^2$Department
  of Chemistry, State University of New York at Stony Brook, Stony
  Brook, New York 11794-3400}
%\end{instit}
\maketitle
%\vfill\eject
\vspace{-0.3cm}
\begin{abstract}
\tighten
  We show the way in which the self-consistent Ornstein-Zernike
  approach (SCOZA) to obtaining structure factors and
  thermodynamics for Hamiltonian models can best be applied to
  two-dimensional systems such as monolayer films. We use the
  nearest-neighbor lattice gas on a square lattice as an
  illustrative example.
\end{abstract}
\narrowtext

\vfill\eject

\section{Introduction}
  The self-consistent Ornstein-Zernike approach (SCOZA) was
  introduced by H\o ye and Stell\cite{scoza1} as an approximation
  method specifically tailored to obtain structure factors and
  thermodynamics for Hamiltonian models in three or more spatial
  dimensions. It was subsequently found by H\o ye and
  Borge\cite{ising2d} that the SCOZA yields extremely accurate
  results for the two-dimensional lattice gas as well, when
  appropriately used, thus opening the way toward the use of SCOZA in
  treating thin-film problems. In this article we summarize the
  two-dimensional SCOZA results. We point out that those results for
  systems considered in the thermodynamic limit are strikingly similar
  to the results that would be found in an exact analysis of
  two-dimensional systems that are finite\cite{mccoy} or
  semi-infinite\cite{onsager}. We note
  why this is to be expected, and using the behavior of the specific
  heat as a criterion, we find the size of the finite and semi-infinite
  systems that yield the best match to the SCOZA results for the
  infinite square lattice.

\section{Background}
  The SCOZA is based on an ansatz used by Ornstein and Zernike\cite{oz},
  which is that the direct correlation function $c({\bf r})$
  introduced by those authors has the range of the pair potential.
  In SCOZA, this ansatz is used along with a core condition that
  guarantees that the two-body distribution function $g({\bf r})$
  must be zero for values of ${\bf r}$ for which the pair
  potential is infinite. In a lattice gas, the core-condition
  implies the exclusion of multiple occupancy of a single lattice
  site or cell; for the equivalent Ising model in which the spin
  variable at site $i$ is +1 or -1, the corresponding condition is
  that $\langle s_i s_j\rangle_{i=j} = 1$, which simply reflects the fact that
  the spin must be pointing either up or down with probability
  one.

  An analysis of the Ornstein-Zernike formalism made by one of the
  authors some time ago\cite{stell} showed that the core condition
  plus the assumption that $c({\bf r})$ is proportional to the pair
  potential implies that for short-ranged potentials, a
  two-dimensional system can not have a critical point at nonzero
  temperature. (In three or more dimensions there is no such
  restriction on criticality.) One knows however, that in two
  dimensions, systems such as the nearest-neighbor lattice gas do
  in fact have a critical point at nonzero $T_c$. Hence SCOZA did
  not initially appear to be a promising method for treating
  arbitrarily large two-dimensional1 systems. Its apparent unsuitability is also
  consistent with the observation\cite{stell,scoza2} that the assumption
  that $c({\bf r})$ has the range of the potential implies that for
  short-ranged potentials, the critical exponent $\eta = 0$, so that
  at a critical point at $T_c \neq 0$, where on expects
  $g({\bf r}) -1 \approx const./r^{d-2+\eta}$, SCOZA would yield
  $g({\bf r}) -1 \approx const./r^{d-2}$. In three dimensions, in
  which $\eta \approx 0.03$, and $d-2+\eta \approx 1.03$, this leads
  to negligible error. But for $d = 2$, in which $\eta = 1/4$, it
  represents the difference between $g({\bf r}) -1 \approx const.
  /r^{1/4}$ and a $g({\bf r}) -1$ that does not appropriately decay
  with increasing $r$.

  However, as we shall see below, in two dimensions, the SCOZA
  results for a square lattice of infinite extent are strikingly
  similar in some respects to exact results for either an $N\times
  N$ lattice or an $N \times \infty$ lattice, with $N \approx 22$.

  It is not hard to understand why the assumption of a $c({\bf
  r})$ for an infinite square lattice yields results that mimic
  exact results for a finite system. For a finite system, $c({\bf
  r})$ is limited in range by the finite boundaries of the system.
  One also knows that in an exact analysis of an infinite
  one-dimensional system with short-ranged potential one finds no
  critical behavior for non-zero temperature. From these
  qualitative statements however, it is not clear what values of
  $N$ in an $N\times N$ or $N\times \infty$ lattice will give rise
  to exact results that most closely match SCOZA results for an
  infinite square lattice. As we shall see in Section IV, when one
  uses the behavior of the specific heat as a criterion, $N$ turns
  out to be around $22$.

\section{Theory}
In the following we consider the two-dimensional square lattice gas,
which is isomorphic to the two-dimensional Ising model.
The potential between particles is
\begin{equation}
  v({\bf r_i}-{\bf r_j}) = \left\{
  \begin{array}{ll}
    +\infty, & \mbox{${\bf r_i} = {\bf r_j}$}\\
    -w, & \mbox{i, j nearest neighbors}\\
    0, & \mbox{otherwise.}
  \end{array}
  \right.
\end{equation}
For convenience $w$ is scaled to be $1$ in our calculations. In
this convention the internal energy per spin for the Ising model
$U$ is related to the internal energy per particle for the lattice
gas $u$ via the following relation:
\be
  U = \rho u +\frac12 q \rho - \frac{1}{8} q \, ,
\ee
where $q$ is the number of nearest neighbors ($q = 4$ for the
square lattice), and $\rho$ is the density for the lattice gas.

SCOZA is based on the enforcement of thermodynamic consistency between
different routes to thermodynamics. This imposes the following relation:
\be
  \frac{\partial (\beta \chi^{-1})}{\partial \beta}
  = \frac{\partial^2 (\rho u)}{\partial \rho^2} \, ,
  \label{consistency}
\ee where $\beta = 1/T$, the inverse temperature, $\rho^{-2} \chi$
is the isothermal compressibility, and $u$ is the internal energy
per particle for the lattice gas. These quantities can be given in
terms of correlation functions through fluctuation theory, which
yields \be
  \beta \chi^{-1}= \frac{1}{\rho (1+\rho \tilde{h}(0))} \, ,
  \label{chi1}
\ee
and through the ensemble average of the Hamiltonian, which yields
\be
  u = -\frac{1}{2}q\rho g_1 = -\frac{1}{2}q\rho (1+h_1) \, ,
  \label{energy}
\ee
where $h({\bf r}) \equiv g({\bf r}) - 1$, and $g({\bf r})$ is the two-body
distribution function. Here $g_1$ and $h_1$ represent the functional values
of $g({\bf r})$ and $h({\bf r})$ at nearest-neighbor positions, and
$\tilde{h}({\bf k})$ is the Fourier transform
of $h({\bf r})$, which is related to the direct correlation function
$c({\bf r})$ by the Ornstein-Zernike equation:
\be
  h({\bf r_i}) = c({\bf r_i}) + \rho \sum_{j} c({\bf r_j})
  h({\bf r_i}-{\bf r_j}) \, ,
  \label{OZ}
\ee
or in the Fourier-transformed space:
\be
  1+\rho \tilde{h}({\bf k}) = \frac{1}{1-\rho \tilde{c}({\bf k})} \, .
  \label{fourier}
\ee

The above relations are exact. In order to proceed we shall approximate
the form of the direct correlation function $c({\bf r})$ by using the
ansatz introduced by Ornstein and Zernike (OZ) that $c({\bf r})$ has
the range of the pair potential. In SCOZA this can be done by generalizing
somewhat the mean spherical approximation (MSA), in which
\be
  \tilde{c}({\bf k}) = c_0 + q c_1 \Phi ({\bf k}) \, ,
  \label{SCOZA}
\ee
where $\Phi ({\bf k})$ is the nearest neighbor sum,
\be
  \Phi ({\bf k}) = \frac{\cos k_x + \cos k_y}{2} \, ,
  \ \ \ \mbox{for a square lattice},
\ee
and $c_0$ and $c_1$ are functions of $(\rho,\beta)$. Eq.~(\ref{SCOZA})
is the OZ ansatz applied to the lattice gas. The MSA is the special case
obtained by setting $c_1 = \beta$ and adjusting $c_0$ to be compatible
with the core condition that assigns zero probability to multiple occupancy
of a single site. In SCOZA one instead adjusts $c_1$ to insure
self consistency between Eq.~(\ref{chi1}) and ~(\ref{energy}).

The core condition $h(0)= -1$ implies a relation between $c_0$ and $c_1$
through the OZ equation:
\bea
  1-\rho &=& \int\!\frac{d^2{\bf k}}{(2\pi)^2} \,\frac{1}
  {1-\rho\tilde{c}({\bf k})} \nonumber \\
  &=& \frac{1}{1-\rho c_0}\int\!\frac{d^2{\bf k}}{(2\pi)^2} \,
  \frac{1}{1-z\Phi({\bf k})} \nonumber \\
  &\equiv & \frac{P(z)}{1-\rho c_0} \, ,
  \label{core}
\eea
where $z \equiv q\rho c_1/(1-\rho c_0)$. $P(z)$ is the value for the
lattice Green function $P(z,{\bf r})$ at ${\bf r} = 0$. For a two-dimensional
square lattice we have
\be
  P(z) = \frac{2}{\pi}\int_{0}^{\frac{\pi}{2}}\frac{d\varphi}{
  \sqrt{1-z^2\sin^2 \varphi}} = \frac{2}{\pi} K(z) \, ,
\ee
where $K(z)$ is the complete elliptic integral of the first kind.
From Eq.~(\ref{core}) we have
\be
  c_0 = \frac{1}{\rho}\left[1-\frac{P(z)}{1-\rho}\right] \, ,
\ee
and
\be
  c_1 = \frac{zP(z)}{q\rho (1-\rho)} \, .
\ee
By taking ${\bf r_i} = 0$ in Eq.~(\ref{OZ}), we get
\be
  -1=h(0)=c_0-\rho c_0 + q\rho c_1 h_1 \, ,
\ee
so
\be
  h_1 = -\frac{1}{q\rho c_1}[1+(1-\rho)c_0] = -\frac{1-\rho}{\rho}
  \frac{1-P(z)}{zP(z)} \, .
\ee

After substitutions Eq.~(\ref{chi1}) and (\ref{energy}) become
\be
  \beta \chi^{-1} = \frac{(1-z)P(z)}{\rho(1-\rho)}
  \label{chi2}
\ee
and
\be
  u = -\frac12 q \left(\rho - (1-\rho)\frac{1-P(z)}{zP(z)} \right) \, .
\ee
From Eq.~(\ref{chi2}) we conclude that in SCOZA the criticality,
if any, occurs at $z = 1$, since at the critical point one has
$\chi^{-1}= 0$,
and we have $P(z) > 0$ for all $z$.

Finally, by applying the thermodynamic consistency via
Eq.~(\ref{consistency}), we get the SCOZA partial differential equation:
\be
  \frac{1}{\rho (1-\rho)}\frac{\partial}{\partial \beta}[(1-z)P(z)]
  = -\frac{q}{2} \frac{\partial^2}{\partial \rho^2} \left[
  \rho(1-\rho)\frac{P(z)-1}{zP(z)}\right] - q \, .
\ee
The boundary conditions are $z=0$, i.e., $P(z)=1$, at $\beta =0$
and $\rho = 0, 1$.

\section{Specific Results and Discussion}
Since for the two-dimensional square lattice $P(z)$ diverges when
$z \rightarrow 1$, the renormalized inverse temperature parameter
$c_1$ also diverges. As a result, SCOZA fails to predict a true
critical point above zero temperature in this case. Nevertheless,
the SCOZA results for $u$ match the exact Onsager
expression\cite{onsager} for $u$ remarkably well over the whole
temperature range. Instead of having an infinite slope at the
exact critical temperature, the SCOZA slope achieves its maximum
at a temperature within a fraction of a percent of the exact
value. The nonsingular but near-singular behavior near the ideal
transition temperature makes our results strikingly similar to the
exact results for a finite-size Ising model on a square lattice,
$N\times N$\cite{mccoy}, or a finite-width strip, $N\times
\infty$\cite{onsager}, for an $N$ a bit greater than 20.

In Fig.~1 we plot the negative internal energy $-U$ versus the
inverse temperature $\beta$ along the critical isochore $\rho =
\rho_c =1/2$, i.e., magnetization being equal to 0, for the
comparison between SCOZA and both infinite- and finite-size Ising
exact results. The comparison is made even clearer by plotting
their residuals in Fig.~2. We find that the deviations between
SCOZA and the other results are very small over the whole
temperature range. The deviations get larger near the critical
point $\beta = \beta_c$, although the largest deviation is still
within 3 percent. Comparing with finite-size and finite-width
exact results, we find this deviation gets minimized when we
choose $N=22$ for an $N\times N$ Ising model or $N=21$ for an
$N\times \infty$ Ising model.

In Fig.~3 we plot the specific heat versus $\beta$ along the
critical isochore. The SCOZA result has a specific heat that stays
finite at its maximum, as all the finite-size and finite-width
specific heats do in an exact theory. In this comparison we again
find that the SCOZA infinite-lattice result has great resemblance
to the $22\times 22$ or $21\times \infty$ Ising model. The SCOZA
curve matches well with the $21\times \infty$ Ising model for
$\beta < \beta_c$. But for $\beta > \beta_c$ the SCOZA result has
somewhat less difference to the $22\times 22$ solution. The
maximum of SCOZA curve occurs at $\beta_{SCOZA} = 1.758$, whereas
the exact critical point for the infinite Ising model is $\beta_c
= 1.763$. The maximum for the $22\times 22$ and $21\times \infty$
Ising models occur at $\beta_{22\times 22} = 1.735$ and
$\beta_{21\times \infty} = 1.764$, respectively. The deviations
between these maximum temperatures and the exact critical
temperature is of the same order for the SCOZA and $21\times
\infty$ results, whereas it is a bit larger for the $22\times 22$
case. In this regard the SCOZA is a bit more similar to the
$21\times \infty$ model.

For each fixed value of $\rho$, we define the temperature
where the specific heat is at its maximum as the pseudo
singular temperature. Here we try to determine a 'pseudo'
spinodal curve by collecting the set of $(\rho, T)$ points that correspond
to those pseudo singularities. This 'pseudo' spinodal curve is
shown in Fig.~4. Note that the points on the curve don't mark
real singularities, and both the specific heat and the compressibility
remain finite, but very large, at these points.
Furthermore, near the pseudo critical point $\rho = \rho_c = 1/2$,
by defining $\Delta \rho = \rho-\rho_c$, $\Delta T=T_c - T$,
we find
\be
\Delta \rho \sim (\Delta T)^{\beta_{spinodal}} \, ,
\ee
with a classical exponent $\beta_{spinodal} = 1/2$ when $T < T_c$.

From Eqs.~(\ref{fourier}) and (\ref{SCOZA}) we find that in SCOZA
the function $1+\rho \tilde{h}({\bf k})$ has a form proportional
to the Fourier transform of the lattice Green function $P(z, {\bf k})$.
Hence we have $\delta ({\bf r}) + \rho h({\bf r}) \propto P(z,{\bf r})$.
From the fact that
\be
  h({\bf r}) \sim P(z,{\bf r}) \sim \exp(-2r\sqrt{\frac{1-z}{z}})
\ee
when ${\bf r} \rightarrow \infty$\cite{montroll}, we find in
SCOZA the correlation length
\be
  \xi = \frac12 \sqrt{\frac{z}{1-z}} \, .
\ee In Fig.~5 we plot the correlation length $\xi$ versus
temperature at the critical isochore. Since for the
two-dimensional SCOZA scheme $z \rightarrow 1$ only when $T
\rightarrow 0$, the correlation length keeps finite at the exact
critical temperature $T_c$ and only diverges when $T\rightarrow
0$. However, when $T=T_c$ we have $\xi \simeq 44$ according to the
SCOZA scheme, which already indicates strong correlations.
Furthermore, the correlation length increases sharply right below
$T=T_c$.

\acknowledgments
  The authors gratefully acknowledge the support of the Division of
  Chemical Sciences, Office of Basic Energy Sciences, Office of
  Energy Research, U. S. Department of Energy.

\begin{figure}[Fig1]

\caption{Negative internal energy for the two-dimensional Ising model $-U$,
  versus the inverse temperature $\beta$.
  The solid curve is the SCOZA result. Other curves are exact solutions for
  $22\times 22, 21\times \infty$, and $\infty \times \infty$ Ising models,
  respectively. $\beta_c$ is the exact critical point for the $\infty \times
  \infty$ model.}
\end{figure}

\begin{figure}[Fig2]

\caption{Residuals between SCOZA and exact results. For optimal
  finite-size results, the deviations are smaller than the $\infty
  \times \infty$ exact results.}
\end{figure}
\begin{figure}[Fig3]

\caption{Constant volume specific heat $C_V$ for SCOZA compared with
  exact results. Note that the specific heat for SCOZA doesn't
  diverge at its maximum, showing a resemblance with finite-size
  models.}
\end{figure}

\begin{figure}[Fig4]

\caption{Pseudo spinodal curve in the $(T,\rho)$ plane, derived from SCOZA.
  Near the pseudo critical point $\rho = 1/2$, there is the relation
  $\Delta \rho \sim (\Delta T)^{\beta_{spinodal}}$ with $\beta_{spinodal} =
  1/2$.}
\end{figure}

\begin{figure}[Fig5]

\caption{Correlation length $\xi$ versus temperature from the SCOZA
  results. Note that $\xi$ starts to increase sharply for $T\simeq T_c$.}
\end{figure}

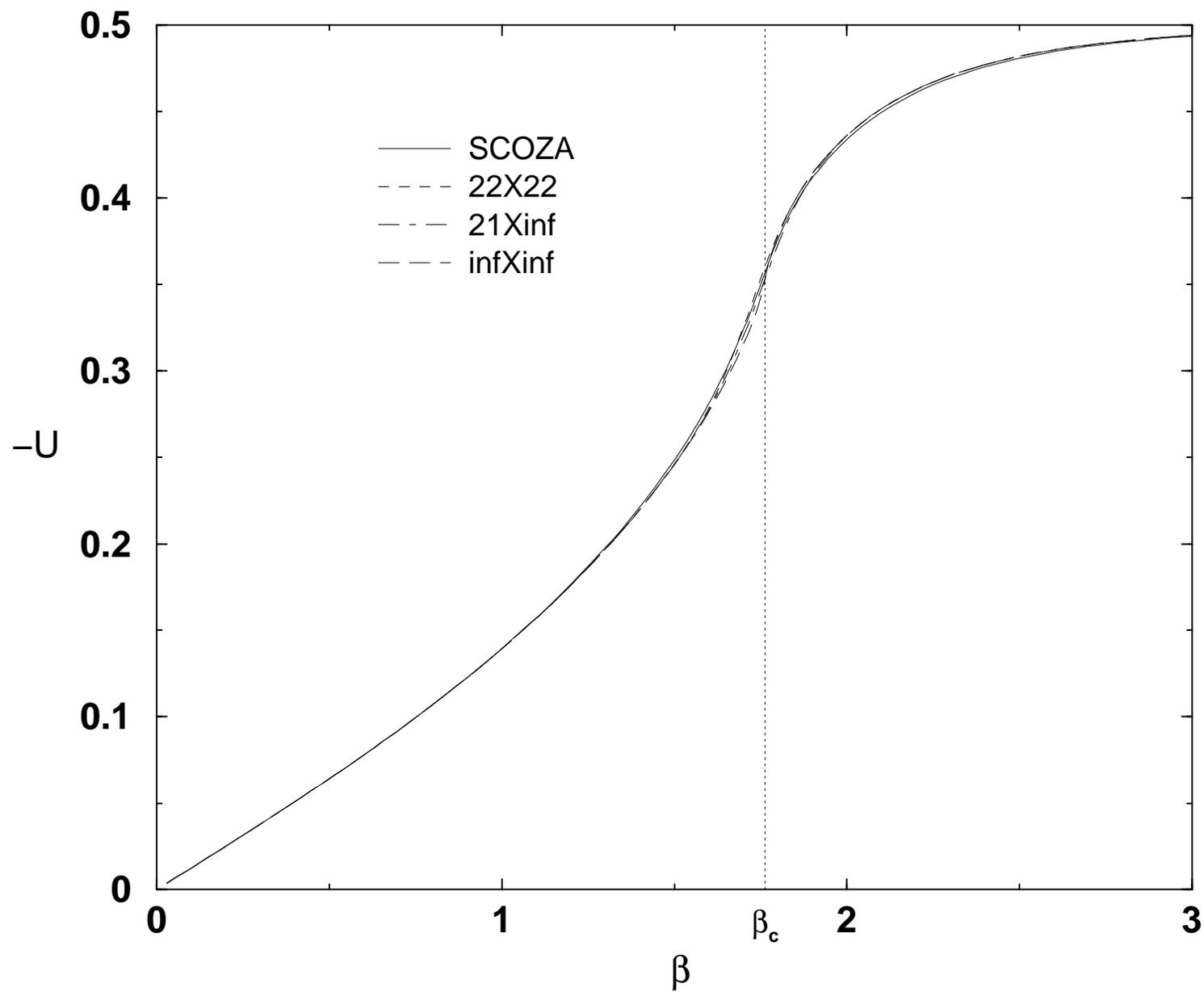
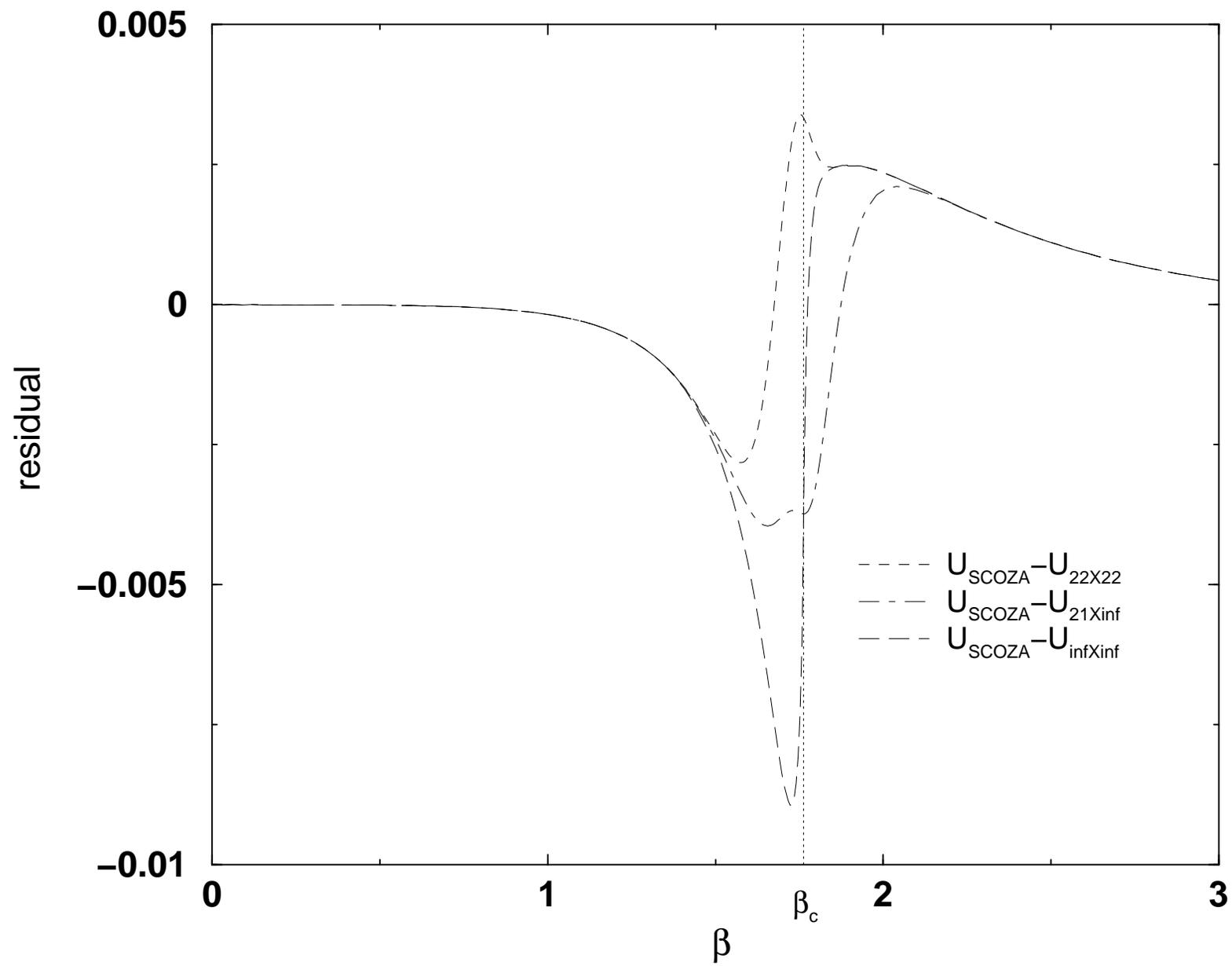
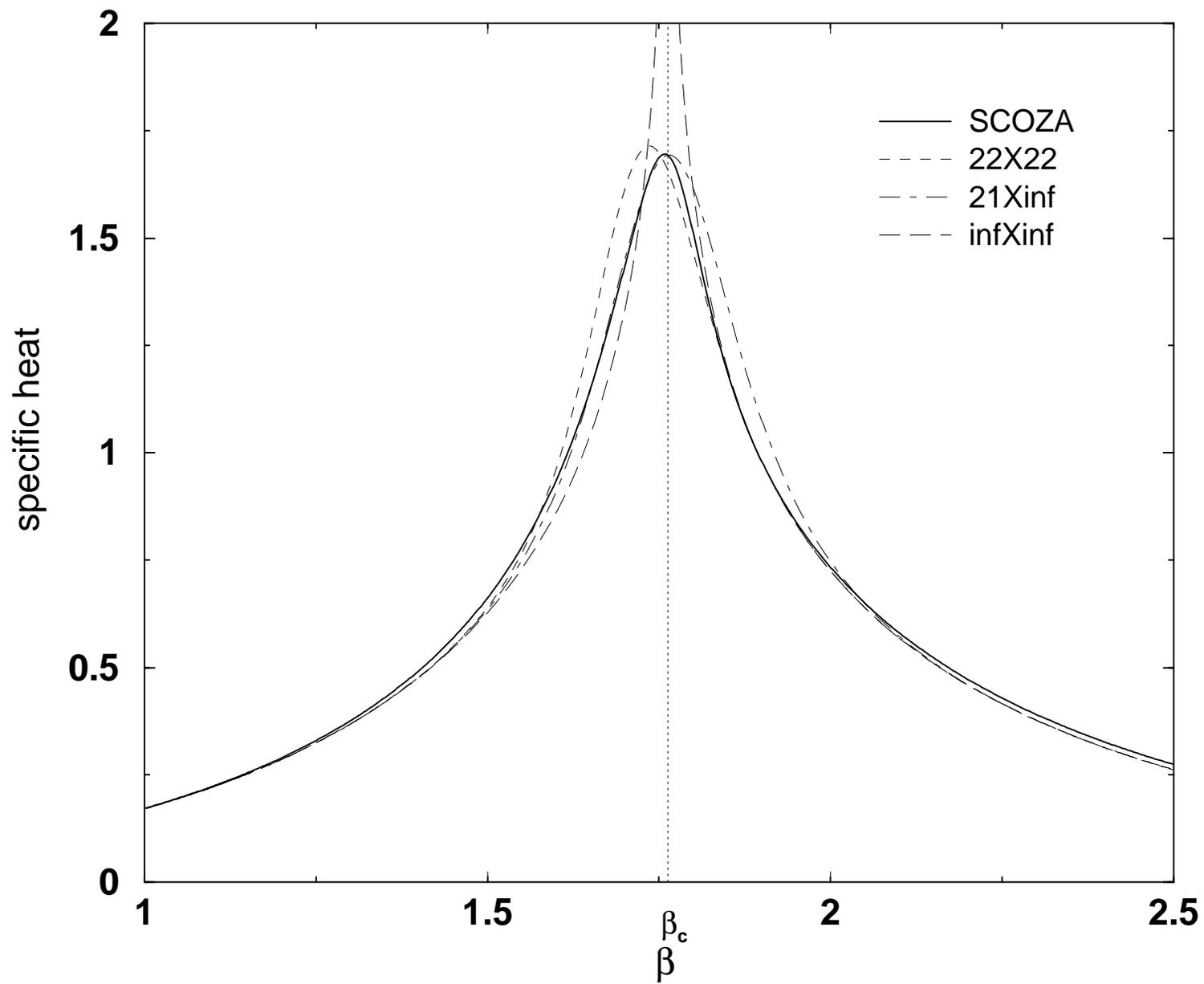
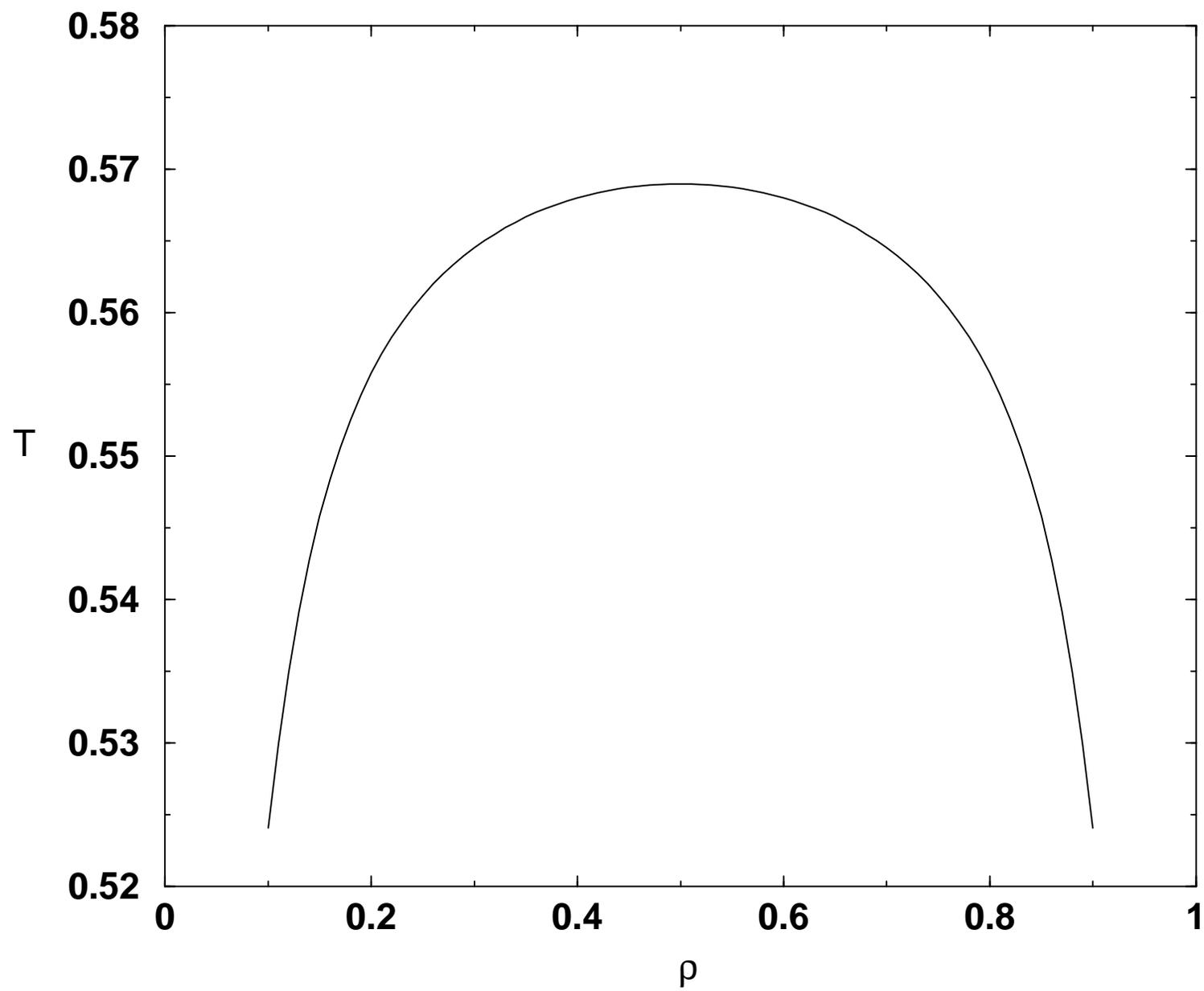
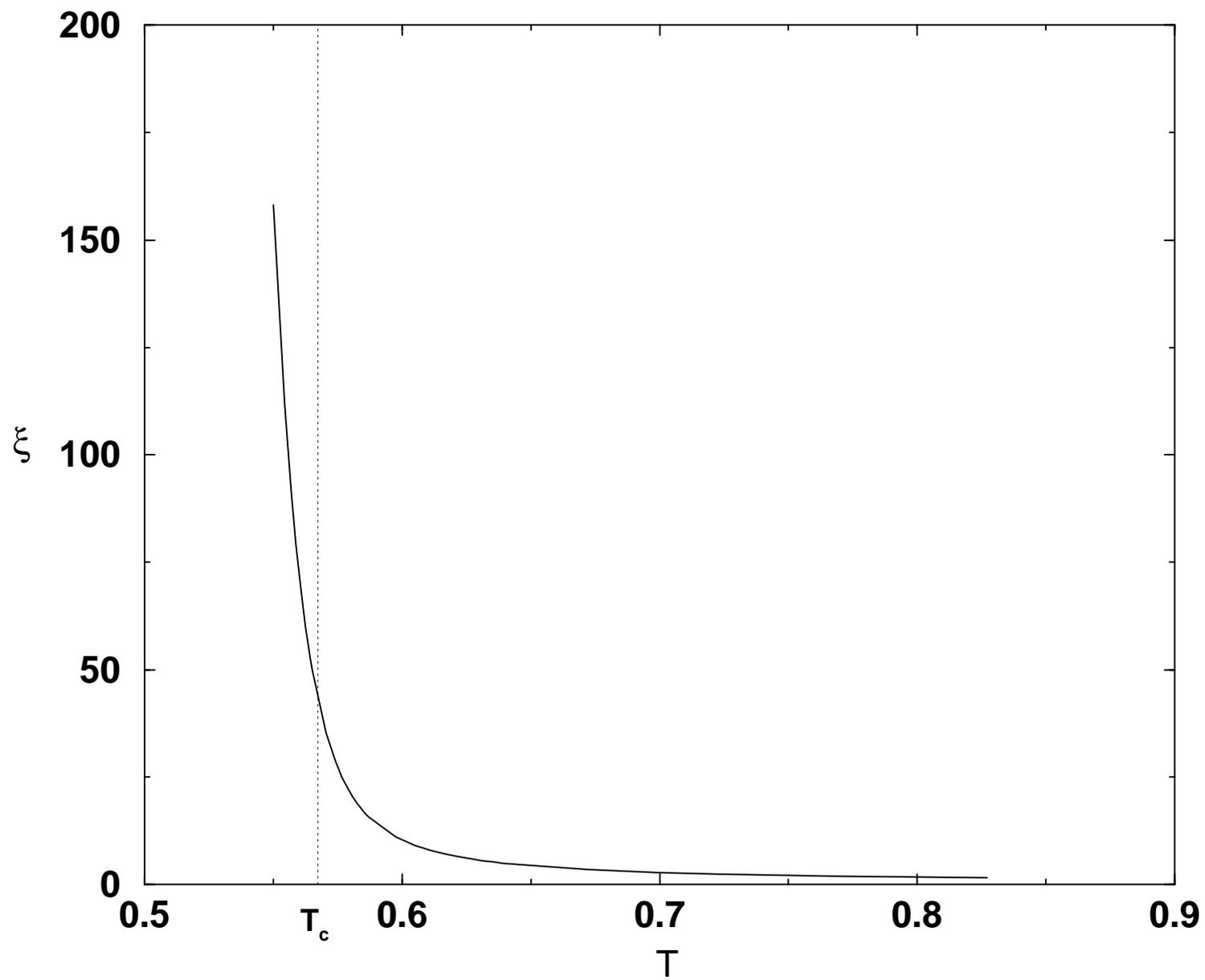
\end{document}